\documentclass[preprint,showpacs,preprintnumbers,amsmath,amssymb]{revtex4}

\usepackage{graphicx}% Include figure files
\usepackage{dcolumn}% Align table columns on decimal point
\usepackage{bm}% bold math

\begin{document}

\preprint{APS/123-QED}

\title{Hitting Time Distributions in Financial Markets}

\author{Davide Valenti$^{\circ}$, Bernardo Spagnolo$^{\star}$ and Giovanni Bonanno}%
 \affiliation{Dipartimento di Fisica e Tecnologie Relative,
  Group of Interdisciplinary
  Physics\footnote {Electronic address: http://gip.dft.unipa.it},
  Universit\`a di Palermo,
 \\
Viale delle Scienze pad.~18, I-90128 Palermo, Italy
\\$^{\circ}$valentid@gip.dft.unipa.it,
$^{\star}$spagnolo@unipa.it}%Lines break automatically or can be forced with \\
\date{\today}% It is always \today, today,

\begin{abstract}
We analyze the hitting time distributions of stock price returns in
different time windows, characterized by different levels of noise
present in the market. The study has been performed on two sets of
data from US markets. The first one is composed by daily price of
1071 stocks trade for the 12-year period 1987-1998, the second one
is composed by high frequency data for 100 stocks for the 4-year
period 1995-1998. We compare the probability distribution obtained
by our empirical analysis with those obtained from different models
for stock market evolution. Specifically by focusing on the
statistical properties of the hitting times to reach a barrier or a
given threshold, we compare the probability density function (PDF)
of three models, namely the geometric Brownian motion, the GARCH
model and the Heston model with that obtained from real market data.
We will present also some results of a generalized Heston model.
\end{abstract}

%Use showkeys class option if keyword
\maketitle

\section{\label{sec:intro}Introduction}
\vskip-0.4cm

The interest of physicists in intedisciplinary researches has been
largely increased in recent years and one of the developing field in
this context is econophysics. It applies and proposes ideas, methods
and models in statistical physics and physics of complex systems to
analyze data coming from economical phenomena \cite{Farmer}. Several
statistical properties verified in financial quantities such as
relative price changes or returns and their standard deviation, have
enabled the establishment of new models which characterize systems
ever better \cite{Mantegna-Stanley}. Moreover the formalism used by
physicists to analyze and to model complex systems constitutes a
specific contribution that physics gives to many other fields.
Complex systems in fact provide a very good paradigm for all those
systems, physical and non-physical ones, whose dynamics is driven by
the nonlinear interaction of many agents in the presence of
"\emph{natural}" randomness~\cite{Anderson}. The simplest
\emph{universal} feature of financial time series, discovered by
Bachelier~\cite{Bachelier}, is the linear growth of the variance of
the return fluctuations with time scale, by considering the relative
price changes uncorrelated. The availability of high frequency data
and deeper statistical analyses invalidated this first approximated
model~\cite{Mantegna-Stanley}, which is not adequate to catch also
other important statistical peculiarities of financial markets,
namely: (i) the non-Gaussian distribution of returns, (ii) the
intermittent and correlated nature of return amplitudes, and (iii)
the \emph{multifractal} scaling~\cite{Borl-Bouc}, that is the
anomalous scaling of higher moments of price changes with time.

In this paper we focus our attention on the statistical properties
of the first hitting time (FHT), which refers to the time to achieve
a given fixed return or a given threshold for prices. Theoretical
and empirical investigations have been done recently on the mean
exit time (MET)~\cite{Mas} and on the waiting times~\cite{Scalas} of
financial time series. We use also the term "\emph{escape time}" to
include the analysis of times between different dynamical regimes in
financial markets done in a generalized Heston
model~\cite{spagnolo-pre06}. Markets indeed present days of normal
activity and extreme days where high price variations are observed,
like \emph{crash} days. To describe these events, a nonlinear
Langevin market model has been proposed in
Refs.~\cite{BouchaudCont}, where different regimes are modelled by
means of an effective metastable potential for price returns with a
potential barrier. We will discuss three different market models
evidencing their limits and features, by comparing the PDF of
hitting times of these models with those obtained from real
financial time series. Moreover we will present some recent results
obtained using a generalized Heston model.

\section{Models for stock market evolution}
\vskip-0.4cm
\subsection{The geometric random walk}
\vskip-0.4cm

The most widespread and simple market model is that proposed by
Black and Scholes to address quantitatively the problem of option
pricing~\cite{Black}. The model assumes that the price obeys the
following multiplicative stochastic differential equation
\vskip-0.7cm
\begin{equation}
  d\,p(t) = \mu \cdot p(t) \cdot dt + \sigma \cdot p(t) \cdot dW(t)
\label{Multiplicative}
\end{equation}
where $\mu$ and $\sigma$ are the expected average growth for the
price and the expected noise intensity (the {\it volatility}) in the
market dynamics respectively. $dp/p$ is usually called \emph{price
return}. The model is a geometric random walk with drift $\mu$ and
diffusion $\sigma$. By applying the Ito's lemma we obtain for the
logarithm of the price
\begin{equation}
  d\,lnp(t) = (\mu - \frac{\sigma^2}{2}) \cdot dt + \sigma \cdot
  dW(t)\;.
\label{Additive}
\end{equation}
This model catches one of the more important stylized facts of
financial markets, that is the short range correlation for price
returns. This characteristic is necessary in order to warrant market
efficiency. Short range correlation indeed yields unpredictability
in the price time series and makes it difficult to set up arbitrage
strategies.

\begin{figure}[htbp]
\includegraphics*[height=7cm,width=11cm]{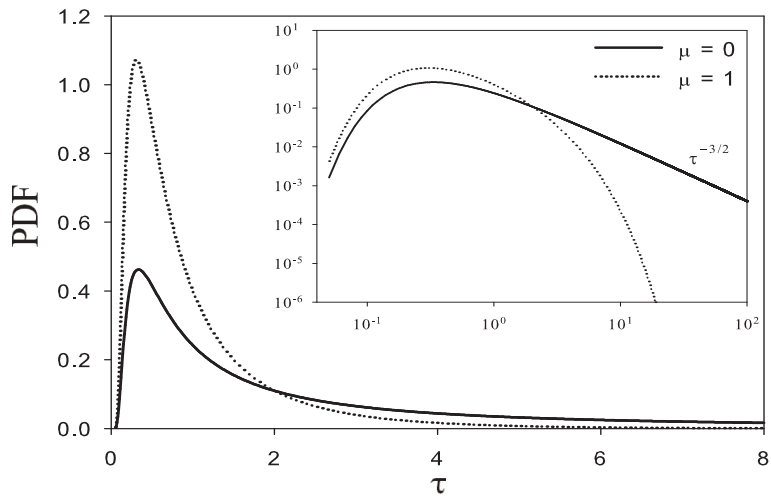}
\vskip-0.5cm
 \caption{Inverse Gaussian distribution obtained with
the parameters $\sigma=1.0$, $(h-p_0) = 1.0$, for $\mu = 0$ (solid
line) and $\mu=1$ (dotted line). Inset: log-log plot of the same
PDF.} \label{F.InvGauss}
\end{figure}

The statistical properties of escape times $\tau$ for this process
are well known, and the PDF of escape time $\tau$, $F(\tau,p_0)$,
was obtained analytically~\cite{Wilmott}. If the starting value of
the price is $p_0$ at time $t=0$, the distribution of the time
$\tau$ to reach a barrier at position $h$ is given by the so called
inverse Gaussian
\begin{equation}
 F(\tau,p_0) = \frac{h-p_0}{\sqrt{2 \pi \sigma^2 \tau^3}}
 \cdot exp\left[-\frac{(h-p_0-\mu \tau)^2}{2 \sigma^2 \tau}\right],
\label{Invgauss}
\end{equation}
which is well known among finance practitioners  to price exotic
options like barrier options~\cite{Wilmott} or to evaluate the
probability for a firm to reach the zero value where it will remain
forever. The shape of the distribution is shown in
Fig.~\ref{F.InvGauss} for two different cases. The asymptotic
expressions of PDF show in one case a power law tail with exponent
$\emph{-1.5}$
\begin{equation}
F(\tau,p_0)\mid_{\tau \rightarrow \infty}\;\;\simeq
\frac{h-p_0}{\sqrt{2 \pi \sigma^2}}\;\;\tau^{-3/2}\;, \;\;\; \mu = 0
\end{equation}
and a dominat exponential behavior in the other case
\begin{equation}
F(\tau,p_0)\mid_{\tau \rightarrow \infty}\;\;\simeq
\frac{h-p_0}{\sqrt{2 \pi \sigma^2}}\;\;\tau^{-3/2}\;
exp\left[-\frac{\mu^2}{2 \sigma^2}\;\; \tau \right], \;\;\; \mu \neq
0.
\end{equation}

The distribution of hitting times for price returns instead is very
simple. In the geometric Brownian motion the returns are
independent, so the probability to observe a value after a certain
barrier is given by the probability that the "particle" doesn't
escape after $n-1$ time steps, multiplied by the escape probability
at the $n^{th}$ step
\begin{equation}
 F_r(\tau) = (1-p) \cdot p^{n-1} = ((1-p) \cdot exp\left[(n-1)\ln p
 \right], \;\;\; n = \frac{\tau}{\Delta t}
 \end{equation}
where $p$ is the probability to observe a return inside the region
defined by the barrier, $\Delta t$ is the observation time step and
$\tau$ is the escape time. So the probability is exponential in
time. The geometric Brownian motion is not adequate to describe
financial markets behavior, because the volatility is considered as
a constant parameter and the PDF of the price is a log-normal
distribution.

\subsection{The GARCH and the Heston models}

Price returns have indeed properties that cannot be reproduced by
the previous simple model: (i) price return distribution has fat
tails; (ii) price returns have short range correlation but the
\emph{volatility} is a stochastic process with long range
correlation~\cite{Man-Sta95}. The degree of variability in time of
the volatility indeed depends not only on the \emph{Fundamentals} of
the firm but also on the market conditions. Volatility is usually
higher during crisis periods and has also an almost deterministic
intra-day pattern during the trading day, being higher near market
opening and closure. So the volatility can be considered as a
stochastic process itself and it is characterized by long range
memory and clustering. More realistic models to reproduce the
dynamics of the volatility have been developed. Here we will present
two of them: the GARCH and the Heston models.

The GARCH(p,q) process (generalized autoregressive conditional
heteroskedasticity), which is essentially a random multiplicative
process, is the generalization of the ARCH  process and combines
linearly the actual return with $p$ previous values of the variance
and $q$ previous values of the square return~\cite{Garch}. The
process is described by the equations
\vskip-0.4cm
\begin{equation}
  \sigma ^2 _t = \alpha_0 + \alpha_1x ^2 _{t-1} + \dots +
  \alpha ^2 _{q} x ^2 _{t-q} +
  \beta_1 \sigma ^2 _{t-1} + \dots + \beta_p \sigma ^2 _{t-p},
  \;\;\;\;\;\;\;
   x_t = \eta _t \cdot \sigma _t,
\label{garch}
\end{equation}
where $\alpha_i$ and $\beta_i$ are parameters that can be estimated
by means of a best fit of real market data, $x_t$ is a stochastic
process representing price returns and is characterized by a
standard deviation $\sigma_t$. The GARCH process has a non-constant
conditional variance but the variance observed on long time period,
called unconditional variance, is instead constant and can be
calculated as a function of the model parameters. It has been
demonstrated that $x_t^2$ of GARCH(1,1) is a Markovian process with
exponential autocorrelation, while the autocovariance of GARCH(p,q)
model is a linear combination of exponential
functions~\cite{Mantegna-Stanley,Garch}. We will consider the
simpler GARCH(1,1) model
\begin{equation}
  \sigma ^2 _t = \alpha_0 + (\alpha_1\eta_{t-1}^2 + \beta_1)\sigma ^2 _{t-1},
  \;\;\;\;\;
  x_t = \eta _t \cdot \sigma _t \;.
   \label{garch11}
\end{equation}
The autocorrelation function of the process $x_t$ is proportional to
a delta function, while the process $x_t^2$ has a correlation
characteristic time equal to $\tau = \mid ln(\alpha_1+\beta_1)
\mid^{-1}$ and the unconditional variance equal to $\sigma ^2 =
\alpha_0 /(1- \alpha_1 - \beta_1)$. So it is possible to fit the
empirical values of these two quantities by adjusting few
parameters. Specifically $\alpha_1$ and $\beta_1$ regulate the
characteristic time scale in the correlation, while $\alpha_0$ can
be adjusted independently to fit the observed unconditional
variance.

In the Heston model~\cite{Heston} the dynamics is described by a
geometric Brownian motion coupled to a second stochastic process for
the variable $v=\sigma^2$. The model equations are
\begin{eqnarray}
  d\,lnp(t) & = & (\mu - \frac{v(t)}{2}) \cdot dt + \sqrt{v(t)} \cdot dW(t),  \\
  dv(t) & = & a(b-v(t)) \cdot dt + c \sqrt{v(t)} \cdot dZ(t) \nonumber,
\label{heston}
\end{eqnarray}
where $W(t)$ and $Z(t)$ are uncorrelated Wiener processes with the
usual statistical properties $<dW_i> \;= 0, \; <dW_i(t) dW_j(t')> \;
= \delta(t-t')~\delta_{i,j}$, but can be also
correlated~\cite{Silva}. The process for $v$ is called
Cox-Ingersoll-Ross~\cite{CIR} and has the following features: (i)
the deterministic solution tends exponentially to the level $b$ at a
rate $a$ ({\it mean reverting process}); (ii) the autocorrelation is
exponential with time scale $\tau=a^{-1}$. Here $c$ is the amplitude
of volatility fluctuations often called the \emph{volatility of
volatility}. Once again we have a model with short range correlation
that mimics the effective long range correlation of the markets
using large values of $\tau$. The process is multiplicative and
values of $v$ can be amplified in few steps producing bursts of
volatility. If the characteristic time is large enough, many steps
will be required to revert the process to the mean level $b$. So the
longer the memory is, the longer the burst will survive. For little
correlation times the process fluctuates uniformly around the mean
level $b$, whereas for large correlation times $v$ presents an
intermittent behavior with alternating activity of burst and calm
periods. The model has been recently investigated by econophysicists
~\cite{Silva} and solved analytically~\cite{Yako}.

\begin{figure}[htbp]
\includegraphics*[height=6cm,width=9cm]{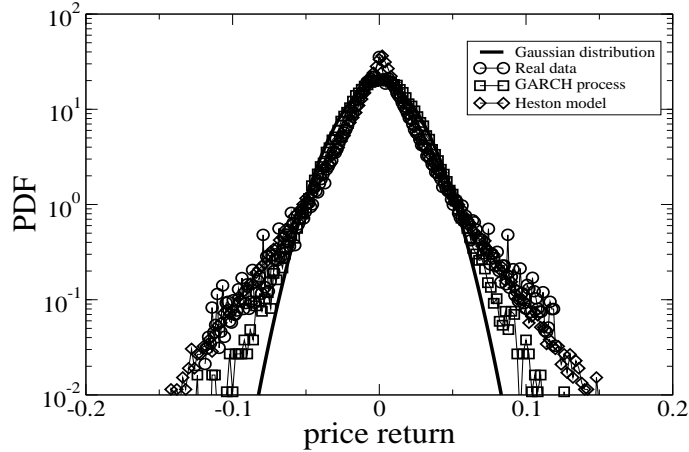}
\vskip-0.5cm
\caption{Probability density function of stock price
returns for: real market data (circle), GARCH model (square) and
Heston model (diamond). The black solid line is a Gaussian
distribution having the same standard deviation of real data.}
\label{F.distro}
\end{figure}

The two models presented so far are a more realistic representation
of financial market than the simple geometric Brownian motion, even
if they do not reproduce quantitatively the form of the long time
correlation observed for the volatility. We use a set of 1071 daily
stock price returns for the 12-year period 1987-1998, and we compare
the results obtained by simulation of the GARCH and Heston models
with those obtained from real market data. The parameters in the
models were chosen by means of a best fit, in order to reproduce the
correlation properties and the variance appropriate for real market.
Specifically for the GARCH model we used values $\alpha_1=0.07906$
and $\beta_1=0.90501$ obtained elsewhere~\cite{Akgiray} to fit the
correlation time of daily market data, and $\alpha_0=7.7 \cdot
10^{-6}$ in order to fit the average standard deviation of our data
using the formula for unconditional variance presented in the
previous section. For the Heston model we used $a=4.5 \cdot
10^{-2}$, $b=8.62 \cdot 10^{-5}$ and $\mu=5.67 \cdot 10^{-4}$,
obtained in a recent work~\cite{Yako}, suitable for daily returns
and $c=10.3 \cdot 10^{-3}$, as before, to fit the average standard
deviation of our data set. Using these parameters we obtain
distribution for price returns that are in reasonable agreement with
real market data, as shown in Fig.~\ref{F.distro}.
The two models approximate the return distributions of real data
better than the Gaussian curve. In particular the Heston model gives
the best agreement. The chosen parameter set therefore is good
enough to fit the dynamics of our data.

\begin{figure}[htbp]
\includegraphics*[height=6cm,width=12cm]{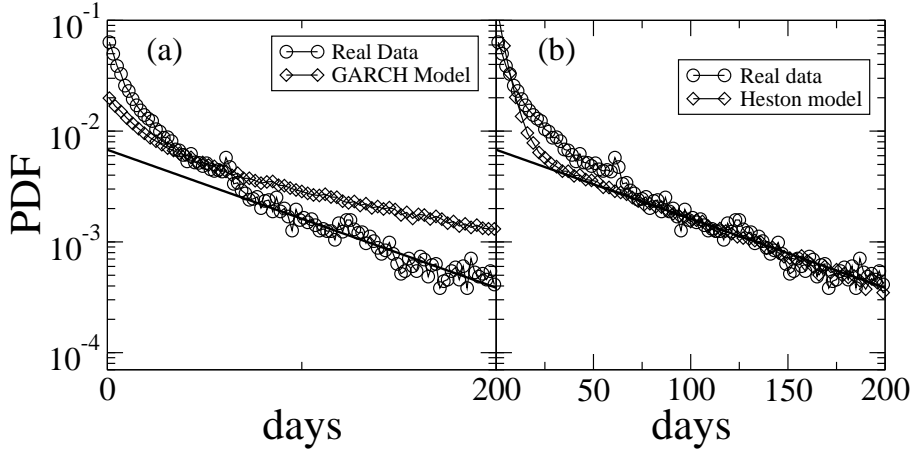}
\caption{PDF of escape times of price returns for the (a) GARCH and
(b) Heston models (diamond) compared with the distribution obtained
from real market data (circle). The process starts at
$(-0.1\sigma)=-0.0022$, the barrier is located at $(-2.0\sigma)$.
The straight line is an exponential fit of the distribution tail.}
\label{F.taudistro}
\end{figure}

In order to investigate the statistical properties of escape times
$\tau$ we choose two thresholds to define the start and the end for
the random walk. Specifically we calculate the standard deviation
$\sigma_i$, with $i=1,\dots,1071$ for each stock over the whole
12-year period. Then we set the initial threshold to the value $0.1
\cdot \sigma_i$ and as final threshold the value $-2 \cdot
\sigma_i$. The thresholds are different for each stock, the final
threshold is considered as an absorbing barrier. The resulting
experimental distribution reported in Fig.~\ref{F.taudistro} has an
exponential tail but it deviates from the exponential behavior in
the region of low escape times. Specifically low escape times have
probability higher than the exponential. We recall that for the
geometric Brownian motion model this distribution should be
exponential over the entire $\tau$ axis. So the first conclusion we
can draw from our analysis is that the basic geometric Brownian
motion is not adequate to explain the distribution of $\tau$.

In order to reproduce more closely the situation present in real
market we choose $\sigma_{start}$ only once, specifically  we place
the random walker in the initial starting position and we set the
initial volatility value. When the random walker hits the barrier,
we register the time and we place the walker again in the initial
position, using the volatility of the barrier hitting time. So the
random walker can experience different initial volatility values as
in real markets. The results are reported in the two panels of
Fig.~\ref{F.taudistro} for both the GARCH and the Heston models, and
we see that these models provide a better agreement with real data
than the geometric Brownian motion. Moreover for the GARCH model the
agreement is only qualitative, whereas the Heston model is able to
fit the empirical distribution quantitatively.

\subsection{The modified Heston model}

Here we consider a generalization of the Heston model, by
considering a cubic nonlinearity. This generalization represents a
fictitious "\emph{Brownian particle}" moving in an \emph{effective}
potential with a metastable state, in order to model those systems
with two different dynamical regimes like financial markets in
normal activity and extreme days~\cite{BouchaudCont}. The equations
of the new model are
\begin{eqnarray}
  dx(t)  &=&  - \left(\frac{\partial U}{\partial x} +
  \frac{v(t)}{2}\right)~dt + \sqrt{v(t)} ~dW_1(t) \\
  dv(t)  &=&  a(b-v(t)) ~dt + c \sqrt{v(t)} ~dW_2(t),
\label{Eqn:BS}
\end{eqnarray}
where $U(x)= 2x^3+3x^2$ is the \emph{effective} cubic potential with
a metastable state at $x_{me} = 0$, a maximum at $x_M = -1$, and a
cross point between the potential and the $x$ axes at $x_I = -1.5$.
In systems with a metastable state like this, the noise can
originate the noise enhanced stability (NES) phenomenon, an
interesting effect that increases, instead of decreasing, the
stability by enhancing the lifetime of the metastable
state~\cite{NES,Nes-theory}. The mean escape time $\tau$ for a
Brownian particle moving throughout a barrier $\Delta U$, with a
noise intensity $v$, is given by the the well known exponential
Kramers law $\tau = exp\left[\Delta U/v \right]$, where $\tau$ is a
monotonically decreasing function of the noise intensity $v$. This
is true only if the random walk starts from initial positions inside
the potential well. When the starting position is chosen in the
instability region $x_o < x_M$, $\tau$ exhibits an enhancement
behavior, with respect to the deterministic escape time, as a
function of $v$. This is the NES effect and it can be explained
considering the barrier "\emph{seen}" by the Brownian particle
starting at the initial position $x_0 $, that is $\Delta U_{in} =
U(x_{max})-U(x_0)$. In fact $\Delta U_{in}$ is smaller than $\Delta
U$ as long as the starting position $x_0$ lies in the interval
$I=[x_I,x_M]$. Therefore for a Brownian particle starting from an
unstable initial position, from a probabilistic point of view, it is
easier to enter into the well than to escape from, once the particle
is entered. So a small amount of noise can increase the lifetime of
the metastable state. For a detailed discussion on this point and
different dynamical regimes see Refs.~\cite{Nes-theory}. When the
noise intensity $v$ is much greater than $\Delta U$, the Kramers
behavior is recovered.

Here, by considering the modified Heston model, characterized by a
stochastic volatility and a nonlinear Langevin equation for the
returns, we study the mean escape time as a function of the model
parameters $a$, $b$ and $c$. In particular we investigate whether it
is possible to observe some kind of nonmonotonic behavior such that
observed for $\tau~vs.~v$ in the NES effect with constant volatility
$v$. We call the enhancement of the mean escape time (MET) $\tau$,
with a nonmonotonic behavior as a function of the model parameters,
NES effect in the broad sense. Our modified Heston model has two
limit regimes, corresponding to the cases $a=0$, with only the noise
term in the equation for the volatility $v(t)$, and $c=0$ with only
the reverting term in the same equation. This last case corresponds
to the usual parametric constant volatility regime. In fact, apart
from an exponential transient, the volatility reaches the asymptotic
value $b$, and the NES effect is observable as a function of $b$. To
this purpose we perform simulations by integrating numerically the
equations~(10) and~(\ref{Eqn:BS}) using a time step $\Delta t=0.01$.
The simulations were performed placing the walker in the initial
positions $x_0$ located in the unstable region $[x_I,x_M]$ and using
an absorbing barrier at $x=-6.0$. When the walker hits the barrier,
the escape time is registered and another simulation is started,
placing the walker at the same starting position $x_0$, but using
the volatility value of the barrier hitting time.

\begin{figure}[htbp]
\vspace{5mm}
\centering{\resizebox{8.5cm}{!}{\includegraphics{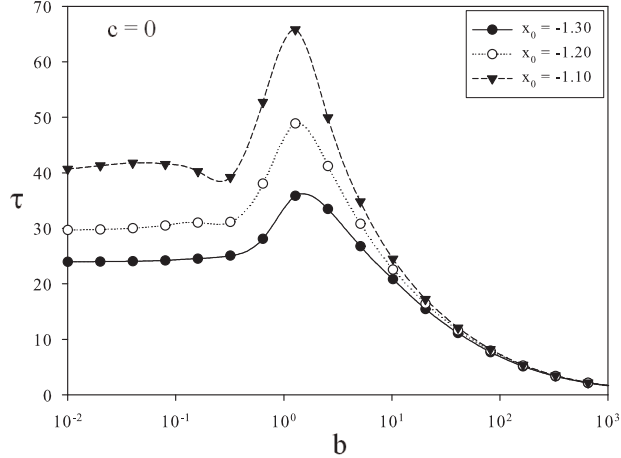}}}
\vskip-0.2cm \caption{\label{limit} Mean escape time $\tau$ for $3$
different unstable starting positions, when only the reverting term
is present: $a = 10^{-2}$, $c = 0$).}
\end{figure}

The mean escape time as a function of $b$ is plotted in
Fig.~\ref{limit} for the $3$ different starting unstable positions
and for $c = 0$. The curves are averaged over $10^5$ escape events.
The nonmonotonic behavior is present. After the maximum, when the
values of $b$ are much greater than the potential barrier height,
the Kramers behavior is recovered. The nonmonotonic behavior is more
evident for starting positions near the maximum of the potential.
For $a = 0$ the system is too noisy and the NES effect is not
observable as a function of parameter $c$. The presence of the
reverting term therefore affects the behavior of $\tau$ in the
domain of the noise term of the volatility and it regulates the
transition from nonmonotonic to monotonic regimes of MET. The
results of our simulations show that the NES effect can be observed
as a function of the volatility reverting level $b$, the effect
being modulated by the parameter $(ab)/c$. The phenomenon disappears
if the noise term is predominant in comparison with the reverting
term. Moreover the effect is no more observable if the parameter $c$
pushes the system towards a too noisy region.  When the noise term
is coupled to the reverting term, we observe the NES effect on the
variable $c$. The effect disappears if $b$ is so high as to saturate
the system.

\begin{figure}[htbp]
\vspace{1mm}
\centering{\resizebox{8.5cm}{!}{\includegraphics{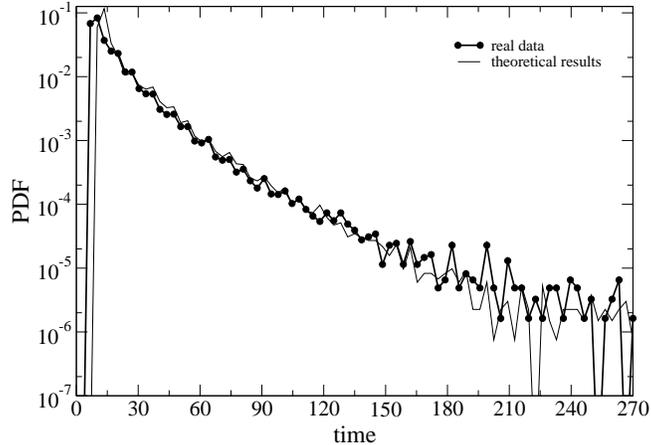}}}
\vskip-0.2cm \caption{Probability density function of the escape
time of the returns from simulation (solid line), and from real data
(black circle).} \label{PDF}
\end{figure}

We compare now the theoretical PDF for the escape time of the
returns with that obtained from the same real market data used in
the previous section. We define two thresholds, $\Delta x_i = 0.1
\sigma_{\Delta x}$ and $\Delta x_f = 1.0 \sigma_{\Delta x}$, which
represent respectively start point and end point for calculating
$\tau$. The standard deviation $\sigma_{\Delta x}$ of the return
series is calculated over a long time period corresponding to that
of real data. The initial position is $x_0 = -1.25$ and the
absorbing barrier is at $x_{abs} = -6.0$. For the CIR stochastic
process $v$ we choose $v_{start}=8.62 \cdot 10^{-5}$, $a=10^{-1}$,
$b=4.5$ and $c=2 \cdot 10^{-1}$. The agreement with real data is
very good. At high escape times the statistical accuracy is worse
because of few data with high values. The parameter values of the
CIR process for which we obtain this good agreement are in the range
in which we observe the nonmonotonic behavior of MET. This means
that in this parameter region we observe a stabilizing effect of the
noise in time windows of prices data for which we have a fixed
variation of returns between $\Delta x_i$ and $\Delta x_f$. This
encourages us to extend our analysis to large amounts of financial
data and to explore other parameter regions of the model.

\section{Escape times for intra-day returns}

\begin{figure}[htbp]
\includegraphics*[height=8cm,width=12cm]{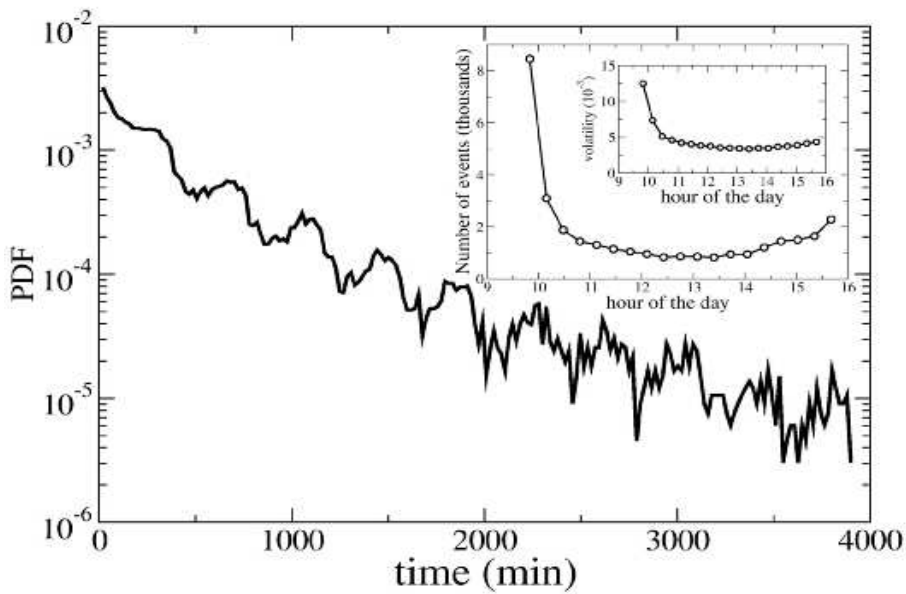}
\vskip-0.5cm
\caption{PDF of escape times of price returns for
intra-day price returns obtained from real market data for the 100
stocks of the S\&P100 basket in the 4--year period 1995--1998. The
process starts at $-0.5\sigma$, the barrier is located at
$-7.0\sigma$. Inset: Frequency distribution of the barrier hitting
event, within a day, as a function of the hour. The internal inset
shows the volatility observed in the same moment of the trading
day.} \label{F.EscapeIntraday}
\end{figure}

In this last section we discuss the results obtained with the same
analysis of the previous section, applied to a different data set at
intra-day time scale. The data set contains 100 stocks in the 4-year
period 1995--1998. The stocks considered are those used, in that
period, in the S\&P100 basket. We are dealing therefore with highly
capitalized firms. The data are extracted from the {\it Trade and
Quote database}. The stocks are distributed in different market
sectors as illustrated in Ref.~\cite{BonannoQF}. For the analysis we
considered the return on a time interval equal to $\delta t=1170
sec$, which is approximately equal to $20$ minute and it is
contained in a market day exactly $20$ times. So we have $19$ price
returns per day, which amounts to $20220$ points in the whole period
of $4$ years per each of the $100$ stocks. We used the value $-0.5
\cdot \sigma$ as a start position and the value $-7.0 \cdot \sigma$
as absorbing barrier. We can choose a so high value for the barrier
because return distribution on intra-day time scale have tails
fatter than daily return distribution, therefore the statistical
accuracy for so high barrier value is good enough. The distribution
of escape times obtained is reported in Fig.~\ref{F.EscapeIntraday}
in a semi-logarithmic plot. It has an exponential trend superimposed
to a fluctuating component. One can recognize that the period of the
fluctuation is 1 trading day, so this effect has to be ascribed to
something that happens inside the daily activity. To describe better
this aspect we record, for each barrier hitting event, the hour when
the event occurs and we build a histogram showing the number of
barrier hitting events as a function of the day time.
This histogram (see the inset of Fig.~\ref{F.EscapeIntraday})
clearly shows that the barrier hitting takes place more frequently
near the opening and the closure of the market. This happens because
the volatility follows a well known deterministic pattern during the
day, being higher near the market opening and closure, and lower in
the middle of the trading day. In the same inset we report an
estimation of the volatility per hour, which we calculate as the
standard deviation of the return observed in that hour, in the whole
period for all the 100 stocks. The figure shows that the volatility
has a pattern reproducing that observed in the barrier hitting event
histogram. This is in agreement with our considerations.

\section{Conclusions}

We studied the statistical properties of the hitting times in
different models for stock market evolution. We discussed
limitations and features of the basic geometric Brownian motion in
comparison with more realistic market models, such as those
developed with a stochastic volatility. Our results indeed show that
to fit well the escape time distribution obtained from market data,
it is necessary to take into account the behavior of market
volatility. In the generalized Heston model the reverting rate $a$
can be used to modulate the intensity of the stabilizing effect of
the noise observed (NES), by varying $b$ and $c$. In this parameter
region the probability density function of the escape times of the
returns fits very well that obtained from the real market data. The
analysis on intra-day time scale shows another peculiarity: the
intra-day volatility pattern produces periodic oscillations in the
escape time distribution. This characteristic will be subject of
further investigation.

\section*{Acknowledgments}
 We gratefully acknowledge Rosario N. Mantegna and the Observatory
 of Complex System that provided us the real market data used for
 our investigation. This work was supported by MIUR and INFM-CNR.

\end{document}